\documentclass[%
 reprint,
 amsmath,amssymb,
 aps,
]{revtex4-2}

\def\mbh{M_{\rm BH}}
\def\rg{r_{\rm g}}
\usepackage{mathtools}

\usepackage{caption}
\usepackage{subcaption}
\def\medd{\dot{M}_{\small{\rm Edd}}}
\newcommand{\mdot}{{\dot{{M}}}}
\newcommand{\rin}{r_{\rm in}}
\newcommand{\rci}{r_{\rm ci}}
\newcommand{\rcm}{r_{\rm cm}}
\newcommand{\rco}{r_{\rm co}}
\newcommand{\rc}{r_{\rm c}}
\newcommand{\vc}{v_{\rm c}}
\newcommand{\thetac}{\Theta_{\rm c}}

\newcommand{\lambdac}{\lambda_{\rm c}}

\newcommand{\vin}{v_{\rm in}}
\newcommand{\thetain}{\Theta_{\rm in}}
\newcommand{\ie}{$i.e.,~$}
\newcommand{\mdotrco}{{\dot{\mathcal{S}}}_{r_{\rm{co}}}}
\newcommand{\mdotrci}{{\dot{\mathcal{S}}}_{r_{\rm{ci}}}}

\def\mbh{M_{\rm BH}}
\def\rg{r_{\rm g}}
\def\rs{r_{\rm s}}
\def\rsh{r_{\rm sh}}
\def\msolar{M_{\odot}}
\newcommand{\qbr}{Q_{\rm br}}

\newcommand{\ndelec}{n_{\rm e^-}}
\newcommand{\ndpos}{n_{\rm e^+}}
\newcommand{\tk}{\tilde{K}}
\newcommand{\ndpro}{n_{\rm p}}
\newcommand{\mpos}{m_{\rm {e^+}}}
\newcommand{\melec}{m_{\rm {e^-}}}
\newcommand{\rhoelec}{\rho_{\rm {e^-}}}
\newcommand{\kb}{k_{\rm {b}}}
\newcommand{\mpro}{m_{\rm p}}
\newcommand{\qvisc}{Q_{\rm visc}}
\newcommand{\trp}{t_{\rm r\phi}}
\newcommand{\as}{a_{\rm s}}
\newcommand{\dq}{\Delta Q}
\newcommand{\trpss}{t_{\rm r\phi,\alpha p}}
\newcommand{\trpr}{t_{\rm r\phi,Real}}
\newcommand{\lambdain}{\lambda_{\rm in}}

\usepackage{graphicx}
\usepackage{dcolumn}
\usepackage{bm}

\newcommand\aap{{A\&A}}

\newcommand\mnras{{Monthly Notices of the Royal Astronomical Society}}
\newcommand\pasj{{PASJ}}

\newcommand\physrep{{PhR}}
\newcommand\araa{Annual Review of Astronomy and Astrophysics}
\newcommand\dps{\ref@jnl{AAS/DPS Meeting Abstracts}}
\newcommand\aj{{AJ}}
\newcommand{\cn}{{\cal N}}
\newcommand{\cd}{{\cal D}}

\usepackage[colorlinks=true, linkcolor=blue, citecolor=blue, urlcolor=blue]{hyperref}
\newcommand{\fullcite}[1]{\citeauthor{#1} (\citeyear{#1}) \cite{#1}}
\usepackage{array,multirow}

\begin{document}


\title{Novel methodology to obtain transonic solutions for dissipative flows\\ around compact objects}

\author{Shilpa Sarkar}
 \email{shilpa.sarkar30@gmail.com}
\affiliation{Harish-Chandra Research Institute, Chattnag Road, Prayagraj,
211019, Uttar Pradesh, India}

\date{\today}

\begin{abstract}
A novel methodology to obtain global transonic solutions around compact objects is reported here. A unified methodology to obtain accretion as well as wind solutions around these objects has been presented. 
Flows around compact objects are dissipative, and the conservation equations are therefore stiff. 
In such conditions, obtaining of sonic point(s) and hence, the transonic solution is not trivial. The conserved equations of motion fail to integrate in the presence of realistic viscosity, thereby making it difficult to obtain a global solution. This inhibits one from getting an actual picture of an astrophysical flow. 
The current work addresses this long-standing issue of obtaining solutions for both accretion and wind. 
The methodology developed utilises the inner boundary conditions and takes recourse to implicit-explicit (ImEx) integration schemes, to obtain general global transonic accretion and wind solutions. 
This is the first time such an attempt has been made. Current work considers the different cooling processes like bremsstrahlung, synchrotron and their inverse-Comptonizations, which are found to affect the thermodynamics of the flow. This methodology could successfully generate all topologies of global solutions, multiple sonic point regime, as well as shocks. 
A broad parameter space study has been done in this work. In an upcoming part II of the paper, a detailed discussion on the spectra and luminosity of the accretion and wind solutions has been presented.

\end{abstract}

\maketitle


\section{\label{sec:1}Introduction}
Accretion onto compact objects is the only mechanism which could explain the extreme luminosities from active-galactic nuclei (AGN, massive BH at the centre), micro-quasars (MQ, stellar-mass BH at the centre) and X-ray binaries (XRBs). The emission is a result of the conversion of gravitational potential energy of the accreted matter into radiation. This accreted matter could directly fall into the central object (Bondi-Hoyle accretion \cite{b52}  or spherical accretion), or form an accretion disc and rotate around the central object for quite a while before being accreted (\citeauthor{ss73} accretion disk \cite{ss73,pr72}). In the presence of strong magnetic fields around compact objects like neutron stars (NSs) and white dwarfs (WDs), the accretion disc is truncated at a certain radius, where the ram pressure of the matter is equal to the magnetic pressure. As soon as the magnetic pressure dominates, matter is channelled along the field lines until it reaches the poles of the object\cite{kol02}. Thus, magnetic fields dictate the dynamics of matter flow.

The current work deals with objects having weak magnetic fields, such that a hydrodynamical treatment is tenable. In these systems, the accretion disc extends down to near the surface of the central object\cite{c96a,bani02,bu19}. 
In the domain of the present study, the topology of accretion flow is similar for all compact objects \ie at infinity they are subsonic, and as the matter is accreted, they become supersonic after passing through a sonic point (SP, hereafter). The transonicity of the flow is mainly due to the strong gravity of the compact objects. 

Apart from accretion, outflows in the form of jets \cite{bz77,bp82} or winds have also been observed in many AGNs, MQ and XRBs. Outflows are just the opposite of accretion\cite{m04}. Matter is ejected at subsonic velocities {and it reaches} supersonic values very far away. \fullcite{c96a}, \fullcite{c96b} connected the methodology to obtain accretion as well as outflow solutions, for different compact objects. The proposition of a unified scheme to obtain transonic solutions was inspired by the works of \fullcite{b52},\fullcite{p58},\fullcite{wd67},\fullcite{ha70},\fullcite{bm76},\fullcite{lt80},\fullcite{fg82},\fullcite{f87} and many more. In all these papers, the location of SP was known apriori. 
In the case of inviscid flows, it is straightforward to obtain the SP conditions\cite{lt80,f87}. The problem arises in the presence of viscosity. At the SP, viscous shear and hence angular momentum ($\lambdac$) is unknown. To skirt out this problem, many works simply assumed shear stress ($\trp$) to be proportional to gas pressure ($\propto p$)\cite{ss73,sc22, sk25} or mixed stress ($\propto p+\rho v^2$, where $\rho$ is the mass density and $v$ is the flow velocity)\cite{cbook90,cm95,c96a,mat84,naka96,sanjit24}. This reduced the azimuthal component of the momentum balance equation to an algebraic form for $\lambda$. Thus, $\lambdac$ is now a function of the radial distance and other flow variables. However, this prescription of stress is over-simplified.
\fullcite{ps94} argued that the viscous stress should be proportional to the change in angular velocity ($\trp\propto d\Omega/dr$, where $\Omega$ is the angular velocity and $r$ is the radial distance) \cite{t07}. Unfortunately, the presence of this gradient term restricts one from obtaining an algebraic expression of $\lambda$, unlike in the earlier cases. Now, to obtain a solution, one needs to solve another differential equation for $\lambda$. This requires 
an additional boundary condition. To skirt out this issue, many authors argued that the viscous shear stress must vanish at the SP \cite{gu00,gp98,pr92,lu99,ct93}. This simplifies the equations of motion and the SP location can be known apriori, reducing the numerical and computational complexity arising from finding SP(s).  However, this boundary condition is arbitrary and stress could only vanish at the horizon\cite{n97,t07}. Keeping in mind all the above drawbacks and utilising the concept of stress-free horizon, \fullcite{bl03}(BL03, hereafter) proposed a methodology to obtain transonic accretion solution using the inner boundary conditions of the BH. Similarly, many authors also used outer boundary conditions \cite{mat84,naka96,c96a,c96b}. However, the outer boundary values assumed were arbitrary. Given a set of constants of motion, a change in outer boundary conditions changes the nature of solutions. Thus, the use of inner boundary conditions are more appropriate, as compared to outer boundary conditions. Most of the information at the inner boundary is known \ie the boundary is asymptotically close to the horizon in case of BH (horizon being a singularity), and the matter should cross it at the speed of light ($c$). The stress-free condition at the horizon further simplifies the conservation equations. All these points were incorporated by BL03. The major drawback of the above work was that, they investigated flows passing through single SPs only.
In relativistic rotating flows,  multiple sonic point (MSP) may form and hence shocks could be present. Shock formation is an important phenomenon, which could explain the observed quasi-periodic oscillations (QPOs) in XRBs and MQs\cite{palit19}. Thus, the origin of MSP is natural and needs to be investigated while modelling of accretion flows and outflows\cite{c97}. 

An additional problem in many models was the use of a fixed adiabatic index ($\Gamma$) equation of state (EoS). The flows around compact objects are trans-relativistic in nature, thus the temperatures vary by orders of magnitude. Also, whether a matter is relativistic or not depends on its mass as well. While electrons become relativistic at much lower temperatures, protons tend to remain non-relativistic or sub-relativistic even at very high temperatures. 
Hence, $\Gamma$ is not constant and would vary w.r.t temperature and mass of the species. \fullcite{cr09} (hereafter, CR) gave an EoS with variable adiabatic index incorporating information of different species. This EoS has an analytical form and hence could be used easily. It accurately fits the relativistically perfect EoS given by \fullcite{c57} {(for ref. see Chapter X, Eq. 236 of the book)}. 

Implementing the CR EoS, and using boundary conditions stated in BL03, many works have scanned the entire parameter space, including the MSP regime\cite{kc14,scp20,sc22,sanjit24,skcp23}. 
However, these works have focused only on accretion flows. Recently, \fullcite{sk25} (Paper 1, hereafter) extended it to obtain wind solutions as well, but for $\trp\propto p$ prescription. This paper elaborately explained an IRM-SP and IRM-SHOCK scheme (Iterative Relaxation Method to obtain Sonic Point; and in few cases finding of SPs through Shocked solutions respectively)  to obtain accretion as well as wind solutions around compact objects. An explicit adaptive Runge-Kutta (RK) method was utilised to solve the conservation equations and obtain solutions. These methods are similar to what has been reported in BL03, but has been extended to obtain MSP and winds. Paper 1 thus provides the latest technique to obtain global transonic solutions, with/without shocks, in the presence of dissipation and also unifies accretion and wind flows. It has a detailed discussion regarding the diagnostics required to obtain SPs and hence a transonic solution.

The work of Paper 1 is extended to incorporate more realistic viscous prescription ($\trp \propto d\Omega/dr$) \cite{kc14,samik23,bl03} in order to investigate trans-relativistic dissipative accretion as well as wind solutions around compact objects.  Apart from proper viscous treatment, this work includes all emission processes relevant in astrophysical systems. They are bremsstrahlung, synchrotron and their inverse-Comptonized components. Magnetic fields are ubiquitous in these flows and hence, stochastic magnetic fields are considered to give rise to synchrotron emission. 
Few of the Compton scatterings might also lead to Compton heating\cite{esin97} of the accretion flow. All these emission mechanisms play an important role in determining the thermodynamics of the flow around compact objects. 
Unfortunately, as has been discussed before, the absence of any algebraic form of the azimuthal component of momentum balance equation forces us to solve an additional differential equation for $\lambda$ with proper boundary conditions (unlike in Paper 1). This equation is not independent but coupled with the other conservation equations. 

\textit{The rise of a major issue! }Solving the conservation equations in the presence of $d\Omega/dr$ is not trivial. This has been reported in Fig. 5 of BL03. When they integrated the equations of motion from the SP (inwards and outwards) to obtain global transonic solutions, they quote that ``...the inwardly directed integration fails". However, integration in outward direction was possible. Such a problem did not arise in Paper 1 of our work, 
where the differential equation for $\lambda$ was absent. 
The issue is indeed serious, since global transonic solutions would always be difficult to obtain. Especially, there is no way one could obtain the wind branch if such an issue persists.
The differential equation of $\lambda$ is stiff because of the presence of $d\Omega/dr$. Stiff equations cannot be solved using the traditional schemes that {have} been employed till now in literature \cite{martin10,martin19,maurya,h99}. Implicit integration methods come to the rescue \cite{l01,but01,nasa}. 

The current work reports the age-old problem of negatively (or inwardly) directed integration of conservation equations in transonic astrophysical flows around compact objects and proposes an implicit-explicit (ImEx) integration scheme to curb the problem \cite{nasa,l01,maurya,but01,t22}. The explicit method is used along with implicit method in order to increase efficiency as well as accuracy. Such an attempt has not been made in the literature. 

The paper is structured as follows: Section 2 discusses the model, equations used and the assumption required to obtain a solution, Section 3 gives an introduction to the issue in detail. Thereafter a discussion on ImEx schemes and the novel methodology to obtain global transonic accretion and wind solutions is presented. Section 4 {analyses} the solutions obtained using the above methodology, and in Section 5, the work is concluded by discussing some implications of the current paper. In an upcoming paper (part II), a detailed study of the spectra and luminosity of the accretion and wind solutions will be presented. 

\section{\label{sec:2}Model and the Mathematical Framework}
In this section, the theory required to model both accretion and wind flows are presented, along with the mathematical formulation used. Although the equations used are applicable to any compact object, focus is particularly made on BHs, where the inner boundary conditions are straightforward. For an accretion flow, it is further simple and matter crosses the horizon supersonically, near the speed of light. 
This work is aimed at addressing the problems that arise while obtaining general viscous, dissipative transonic solutions. So a simplified but robust gravitational potential is assumed, which is the \fullcite{pw80} potential (hereafter, PW), just to avoid getting involved into the complexities of GR. Although simple, this potential mimics the effects of strong gravity. It is to note that, use of different potential or the spin value of the BH does not solve the issue discussed before. 
Steady, axisymmetric advective flow is assumed and a scaling system where, $2G=\mbh=c=1$  is adopted (where $G$ is the Gravitational constant, $\mbh$ is the mass of BH and $c$ is the speed of light in vacuum), such that velocity, length and time are in units of $c$, $\rs=2G\mbh/c^2$ and  $2G\mbh/c^3$ respectively.

\subsection{Equation of State}
Before discussing the conservation equations for different types of flows, the EoS which will be used throughout this work is presented here. An EoS serves as a closure equation and relates the primitive variables like pressure, density and temperature. As discussed in the introduction, the current work uses the CR EoS \citep{cr09} for multi-species flow with variable adiabatic index ($\Gamma$). The expression for energy density of CR EoS is:

\begin{equation}
e=\sum_{i} e_i = \sum_{i} \left[ n_i m_i c^2 + p_i \left( \frac{9p_i + 3n_i m_i c^2}{3p_i + 2n_i m_i c^2} \right)\right]=\dfrac{\rho f}{\tk},
\label{eq:eos}
\end{equation}
where,
\begin{eqnarray}
f&= (2-\xi) \left[ 1+\Theta \left( \dfrac{9 \Theta +3}{3 \Theta +2}\right)\right]+\xi \left[ \frac{1}{\chi}+\Theta \left( \dfrac{9 \Theta +3/\chi}{3 \Theta +2/\chi}\right)\right]\nonumber\\
&{\rm and,~}\tilde{K}=2-\xi(1-1/\chi).\nonumber
\end{eqnarray}
In Eq.~\ref{eq:eos}, the summation implies the sum over $i^{\rm th}$ species, where the species could be either electrons ($e^-$), protons ($p$) or positrons ($e^+$) or a combination of them and $\rho$ is the total mass density. Definitions of other variables are: $\xi=\ndpro/\ndelec$, which is also known as the composition parameter, $\chi=\melec/\mpro$, and $\Theta=\kb T/(\melec c^2)$, which is the dimensionless temperature defined w.r.t the rest mass energy of the electron. Here, $T$ is the temperature in Kelvin and $\kb$ is the Boltzmann constant. The adiabatic index can then be defined as $\Gamma=1+1/N$, where $N=(1/2)~df/d\Theta$ is the polytropic index. Charge neutrality is assumed in the system, implying that the number of positive charges is equal to the number of negative charges. Also, the system is of one-temperature ($T$) \ie all the species have settled down into a single temperature distribution. This assumption relieves us from the complexities present in two-temperature solutions \citep{sc19a, scp20, sc22, skcp23}.

Using the above-discussed relations, the total number density, mass density and gas pressure can be defined in the following way:
\begin{eqnarray}
n&=&\ndelec+\ndpos+\ndpro=2\ndelec, \label{eq:rel1}\\
\rho &=& \ndelec \melec +\ndpos \mpos +\ndpro \mpro =\rhoelec \tk,\\
p&=& (\ndelec + \ndpos +\ndpro)\kb T = \cfrac{2\rho \Theta}{\tk}\label{eq:rel3}.
\end{eqnarray}

\subsection{Conservation equations for different types of flows}
In this section, the conservation equations or the equations of motion (EoM) required to describe the flows around compact objects are discussed. 
The final form of conservation equations are obtained through simplification using the CR EoS and the relations given by Eqs.~\ref{eq:rel1}--\ref{eq:rel3}.

The PW potential has the following form:
\begin{equation}
\Phi=-\frac{1}{r-1},
\end{equation}
where, $r$ is in units of $\rs$. 
The presence of viscosity causes the transport of angular momentum. Thus, the azimuthal component of the momentum balance equation needs to be solved, apart from the equations that have been discussed in Paper 1, which are radial momentum equation, continuity equation and the first law of thermodynamics. Presented below in detail are all the conservation equations required for modelling of flows.

\begin{itemize}
\item Mass accretion rate is given by:
\begin{equation}
\mdot=2\pi \rho v Hr=\pi \Sigma v r.
\label{eq:mdotdisc}
\end{equation}
Here, $H$ is the half-height of the flow and $\Sigma=2H\rho$ is the height-integrated density.  Assuming hydrostatic equilibrium along the transverse direction, the expression for $H$ is \cite{kc14}:
\begin{equation}
H=2\sqrt{\frac{\Theta r}{K}}(r-1).
\label{eq:hh}
\end{equation}
\item The azimuthal component of the momentum balance equation is:
\begin{equation}
\frac{d\lambda}{dr}+\frac{1}{\Sigma v r}\frac{d(r^2 \trp)}{dr}=0,
\label{eq:lambdaeq1}
\end{equation}
where, $\trp$ is the viscous stress. Integrating the above equation using Eq.~\ref{eq:mdotdisc}:
\begin{equation}
\dot{M}(\lambda-\lambda_0)=-2\pi r^2 \trp,
\label{eq:lambdaeq2}
\end{equation}
where, $\lambda_0$ is the specific angular momentum at the horizon. To obtain the above expression, vanishing of the viscous stress at the horizon has been considered \citep{bl03}. 
Discussed below are the two different prescriptions for the $\trp$ term as have been mentioned in the introduction section. 

\begin{enumerate}
\item  First is the general, $\trp \propto \alpha p$  prescription, where $\alpha$ is the \citeauthor{ss73} viscosity parameter \cite{ss73}. This is famously used by many authors, because it reduces Eqs.~\ref{eq:lambdaeq1}--\ref{eq:lambdaeq2} to an algebraic form \citep[Paper 1]{sc22, s09}. The exact expression is given below:
\begin{equation}
\trpss=-\alpha W,
\end{equation}
where, $W$ is the height-integrated gas pressure $W=2Hp$. Then, Eq.~\ref{eq:lambdaeq2} reduces to:
\begin{equation}
\lambda=\lambda_0+\frac{2\pi r^2 \alpha W}{\dot{M}}.
\end{equation}
Therefore, one obtains a simple analytical expression for $\lambda$ as a function of $r$ and $p$. This prescription has been followed in Paper 1.

\item The second prescription is more realistic and involves the gradient of angular velocity \citep{sanjit24,kc14,t07,bl03}. It is given as,
\begin{equation}
\trpr=\eta r \frac{d\Omega}{dr},
\end{equation}
where, $\mathbf{\Omega}$ {is the angular velocity}, $\eta$ is the dynamical viscosity parameter and is given by $\eta=\Sigma \nu=2\rho H [\alpha \as^2/(\Gamma \Omega_{\rm K})]$, where $\nu$ is the kinematic viscosity parameter, $\Omega_{\rm K}=1/[2r(r-1)^2]$ is the Keplerian angular velocity and $\as=2\Gamma \Theta/\tilde{K}$ is the sound speed.
Substituting this expression in Eq.~\ref{eq:lambdaeq2}:
\begin{equation}
\frac{d\Omega}{dr}=-\frac{\Gamma v \Omega_{\rm K}(\lambda-\lambda_0)}{\alpha \as^2 r^2}.
\label{eq:domegadr}
\end{equation}  
Since, $\mathbf{\lambda=r^2\Omega}$, the differential equation for angular momentum is given by:
\begin{equation}
\frac{d\lambda}{dr}=2r\Omega+r^2\frac{d\Omega}{dr}.
\label{eq:dlamdr}
\end{equation}
Thus, one needs to solve the differential Eq.~\ref{eq:dlamdr}, with the help of differential Eq.~\ref{eq:domegadr} to obtain angular momentum at each radial coordinate.
\end{enumerate}

\item The form of first law of thermodynamics is unchanged and is similar to what has been used in Paper 1, except for the inclusion of a different viscous dissipation term inside the flow \citep{kc14} and different cooling mechanisms. Viscous dissipation is known to heat up the system. 
The first law of thermodynamics is given by:
\begin{equation}
~~~~~~~~~\frac{d\Theta}{dr}=-\frac{2\Theta}{2N+1}\left[ \frac{1}{v}\frac{dv}{dr}+\frac{5r-3}{2r(r-1)}\right]-\frac{\dq K}{\rho v (2N+1)},
\label{eq:flt}
\end{equation}
where, $\Delta Q=Q^+-Q^-$. 
Here, $Q^+=Q_{\rm visc}+Q_{\rm compheat}$ and $Q^-=\qbr+Q_{\rm syn}+Q_{\rm bcomp}+Q_{\rm scomp}$, where $Q_{\rm compheat}$ is the Compton heating part while $\qbr$ and $Q_{\rm sync}$ are the cooling due to bremsstrahlung and synchrotron radiation respectively. $Q_{\rm bcomp}$ and $Q_{\rm scomp}$ are the Comptonizations of the bremsstrahlung and synchrotron soft photons respectively. The expressions for each of these cooling processes are given in \fullcite{scp20,sc22}.  The expression for the heating due to $\alpha p$ prescription is given in \fullcite{mat84} as well as Paper 1 of our work. For the realistic prescription, $\qvisc$ expression is taken from BL03.

\item The radial component of momentum balance equation is given by:
\begin{equation}
v\frac{dv}{dr}+\frac{1}{\rho}\frac{dp}{dr}-\frac{\lambda^2}{r^3}+\frac{1}{2(r-1)^2}=0.
\label{eq:dvdro}
\end{equation}
The differential equation for velocity is obtained by simplifying the above equation using Eqs.~\ref{eq:eos}, \ref{eq:mdotdisc}, \ref{eq:hh} and \ref{eq:flt} and is given by:
\begin{equation}
\dfrac{dv}{dr}=\dfrac{\as^2\left[\dfrac{2N}{2N+1} \dfrac{5r-3}{2r(r-1)}\right]+\dfrac{\lambda^2}{r^3}-\dfrac{1}{2(r-1)^2}+\dfrac{\dq}{\rho v (2N+1)}}{v \left[1 -\dfrac{\as^2}{v^2}\left( \dfrac{2N}{2N+1}\right)\right]}.
\label{eq:dvdr1}
\end{equation}
This can be written in the following numerator by denominator form :
\begin{equation}
\dfrac{dv}{dr}=\cfrac{\cal N}{\cal D}.
\label{eq:dvdr2}
\end{equation}

\item Integrating Eq.~\ref{eq:dvdr1}, the generalised Bernoulli constant for a dissipative flow  is obtained \citep{bl03}:
\begin{equation}
\dfrac{1}{2}v^2+h-\dfrac{\lambda^2}{2r^2}+\dfrac{\lambda \lambda_0}{r^2}-\dfrac{1}{2(r-1)}+{\cal Q}={E},
\label{eq:bernoullidisc}
\end{equation}
where, ${\cal Q}=\int (Q_{\rm compheat}-Q^-)dr /(\rho v) $ and $h~=~(e+p)/\rho$ is the enthalpy. In the absence of cooling, the above equation can be simplified to,
\begin{equation}
\dfrac{1}{2}v^2+h-\dfrac{\lambda^2}{2r^2}+\dfrac{\lambda \lambda_0}{r^2}-\dfrac{1}{2(r-1)}={\mathbb{ E}}.
\label{eq:gse}
\end{equation}
This equation is called the grand specific energy \citep{kc14}.
\item {The rate of specific entropy generation} is given by:
\begin{equation}
\dot{\mathcal{S}}=v H r \exp \left(k_3\right) \Theta^{3 / 2}(3 \Theta+2)^{k_1}(3 \Theta+2 / \eta)^{k_2},
\label{eq:entropy}
\end{equation}
where, $k_1=\cfrac{3(2-\xi)}{4}, k_2=\cfrac{3 \xi }{ 4}$ and $k_3=\cfrac{f-~\tk}{2 \Theta}$. The above expression is taken from \fullcite{ketal13}, where they have given a detailed derivation of the same. This formula measures the value of entropy at any given $r$. It is important to note that nature prefers a solution with maximum entropy \cite{b52,scp20}. This expression is mainly used to select solutions with higher entropy in the MSP regime where there are more than one SPs. A flow would prefer that solution which has a higher entropy. 
\end{itemize}

\subsection{Multiple sonic point regime}
SPs are formed where the fluid velocity crosses the effective speed of sound in the flow. 
The presence of gravity, relativistic effect and rotation gives rise to multiple sonic points\cite{f87} or critical points (CP).  It is to note that in the current work, the terminology CP and SP have been used analogously. The finding of SP signifies finding of CP only \cite{suso22, kalyan25}. Thus, from now on the term SP will be used throughout the paper. 

It is at the SP that the $dv/dr$ has a $0/0$ form, or the numerator and denominator goes to ${\cal N}={\cal D}=0$.
The SPs are named according to their distance from the central object: inner ($\rci$), middle ($\rcm$) and outer ($\rco$). $\rci$ and $\rco$ are X-type SPs, \ie matter can actually pass through them, while $\rcm$ is an O-type/spiral-type SP 
and hence, matter do not physically pass through it. Depending on the set of flow parameters or constants of motion supplied, the flow may harbour one or MSP. 
\subsection{Shocks}
\label{sec:shocks}
In the MSP regime, where 2 X-type SPs are present, shock could be formed. In the presence of shocks, for accretion flows: subsonic matter starting from infinity would first pass through $\rco$ and become supersonic 
$\rightarrow$ encounter a shock transition at $\rsh$ $\rightarrow$ become subsonic $\rightarrow$ again become supersonic, after passing through $\rci$. For the case of winds, the subsonic flow will start from near the surface and follow the reverse path: $\rci\rightarrow\rsh\rightarrow\rco\rightarrow\infty$.  
Location of shock is determined from Rankine-Hugoniot (RH) conditions, \ie mass flux, energy flux and azimuthal as well as radial component of momentum flux conservation\cite{ll59}. The conditions are respectively: (i) $\mdot_+=\mdot_-$, (ii) $\dot{E}_+=\dot{E}_-$, (iii)  $\dot{J}_+=\dot{J}_-$ and (iv) $W_++\Sigma_+v_+^2=W_-+\Sigma_-v_-^2$. Here, $-$ and $+$ denote the quantities before and after the shock transition and $\dot{J}=\mdot\lambda+r^2\trp=\mdot \lambda_0$.

It is important to remember that in the MSP region, a shock is preferred by nature since a higher entropy branch solution exists. In accretion flows, the condition reads as: $\mdotrci > \mdotrco$, \ie the entropy accretion rate of $\rci$ should be greater than that of $\rco$ \cite{f87,lt80}. The values of entropy are computed using Eq.~\ref{eq:entropy}. The shock jump is favoured only when the post-shock solution has a higher entropy. After the entropy conditions are checked, RH conditions are used to locate the shock. In case of winds, since matter starts from the surface, the entropy condition is, $\mdotrco > \mdotrci$, which is just the reverse of accretion flows.

Flows where RH conditions are not satisfied in the MSP regime (\ie in the absence of shocks), the flow would pass through a single SP, either $\rci$ or $\rco$. 
The selection of SP and hence, the corresponding transonic solution depends on two factors: (i) solution needs to be global, \ie connects the outer boundary to the inner boundary, and (ii) if both the SPs produce a global solution,n then the solution with higher entropy is preferred by nature.

\section{Methodology}
There have been different methodologies developed so far to obtain transonic solutions. In general, the following two steps are executed: (1) locate the SP or SPs and (2) integrate the EoM from the SP in different directions to obtain the desired accretion/wind solutions. Details about both these steps are given below:
\begin{itemize}
\item Locating SP (Step 1): As have been discussed in the introduction, methodology used to locate SP differ for flows with dissipation and without dissipation and is not trivial for the latter case. 
In the current work, the IRM-SP and IRM-SHOCK scheme of Paper 1 is utilised. IRM-SP is an extended version of the work done by BL03. This is a general scheme and can be employed irrespective of whether the system is dissipative or not. It utilises the inner boundary conditions to locate the SP. 
\begin{itemize}
\item IRM-SP scheme is used to locate all SPs of accretion and wind flows which are connected to the inner boundary. 
\item IRM-SHOCK: This method is used only to find a certain class of wind solutions which harbour shocks. In MSP regime for wind flows, the solution passing through outer SP ($\rco$) is sometimes non-global, \ie not connected from infinity to the inner boundary. 
In that case, the solution passing through $\rci$ is global. This is because both solutions cannot be non-global, and a physical global solution should exist for every constant of motion. IRM-SHOCK is used to find $\rco$ just for this specific case. 

There could be two possibilities for the wind flow for the above case: (i) flow could pass through $\rci$, since this solution is global, connects the surface of the compact object to infinity or (ii) the solution passing through $\rci$ might encounter a shock transition and then would flow through $\rco$. The latter case allows one to locate $\rco$ although it is non-global and not connected to the inner boundary. Locating $\rco$ (which has a non-global solution) is important only in cases when there is a shock, thus the name IRM-SHOCK. Otherwise, the flow would pass through a single SP ($\rci$), which is connected to the inner boundary and could be located using IRM-SP methodology solely (as mentioned in point i). 
IRM-SHOCK methodology was thus developed to locate the outer SP of wind solutions, through shock transition (see, Paper 1 for more details). 
\end{itemize}

The above methodology of finding SPs, for accretion as well as winds, has been well established in Paper 1. In the current paper, these methodologies will be used exactly.

\item Finding solution (Step 2):
The major problem arises only after the SP has been located, and one proceeds to obtain a solution for accretion/wind by integrating the EoMs. The current work deals with this issue. 
 In the absence of viscosity or in the presence of $\alpha p$ viscosity prescription (Paper 1) the inwardly-directed integration from SP ($dr=-h$, where $h$ is the step size) to obtain global TS is possible, as has been seen in Paper 1. However, BL03 pointed out that this inwardly-directed integration is not possible for flows with realistic viscous shear stress prescription, \ie $\trp\propto d\Omega/dr$. 
\end{itemize}

 In this section,  the issue is elaborately explained and a methodology is proposed to obtain global TS. For appropriateness,  focussed are only the solutions that are connected to the inner boundary. Thus, IRM-SP scheme is sufficient to locate the SPs. The discussion is divided into two subsections: 
\begin{enumerate}
\item LOCATE SP(s): Revisit the IRM-SP scheme, where a short explanation on the method to obtain the SP in the presence of realistic viscosity prescription is presented.
\item OBTAIN TSs: Discuss the methodology to obtain the global TSs for accretion and winds. This part is subdivided into two further subsections.
\begin{itemize}
\item One section discusses the issues associated {with} using the traditional methods of integration, to obtain global TS.
\item   Second part {discusses} the scheme which can remove this issue and allow one to obtain global TSs for accretion as well as winds. 
\end{itemize}
\end{enumerate}

\subsection{Method to locate SPs in presence of $\trpr$ viscosity: Extension of IRM-SP to IRM-SP-RV}
The IRM-SP scheme of Paper 1 uses the explicit adaptive Runge-Kutta (RK) method to solve the conservation equations to find SP as well as the TS. Traditionally, explicit methods are used for integration in steady-state cases. These methods are fast as well as efficient. Paper 1 used the Cash-Karp method, which is a fifth-order accurate explicit integration scheme\cite{ck90}. This higher-order method has higher stability than its lower-order counterparts, and could effectively adapt its step size depending on the velocity gradients. This method could successfully solve the coupled ordinary differential equations with steep slopes. However, in the current work, the presence of $\trpr$ viscosity prescription requires one to solve an additional differential equation for angular momentum. Below, the IRM-SP methodology has been extended to include $\trpr$ viscosity and is named as IRM-SP-RV (where, RV stands for Realistic Viscosity prescription). 

Both these methods are almost the same except that an additional equation, which is the azimuthal component of angular momentum equation, needs to be solved. It is important to note that additional cooling processes like synchrotron, Comptonization of soft synchrotron and bremsstrahlung photons and Compton heating, have been included, which were not considered in Paper 1. The inclusion of these cooling processes does not alter the methodology of finding solutions, but are included as additional algebraic terms in the first law of thermodynamics (see Eq. \ref{eq:flt}). The IRM-SP-RV method has been described below. 

\begin{figure}
\includegraphics[scale=0.78,  trim={3.6cm 2.cm 10cm 6.cm},clip]{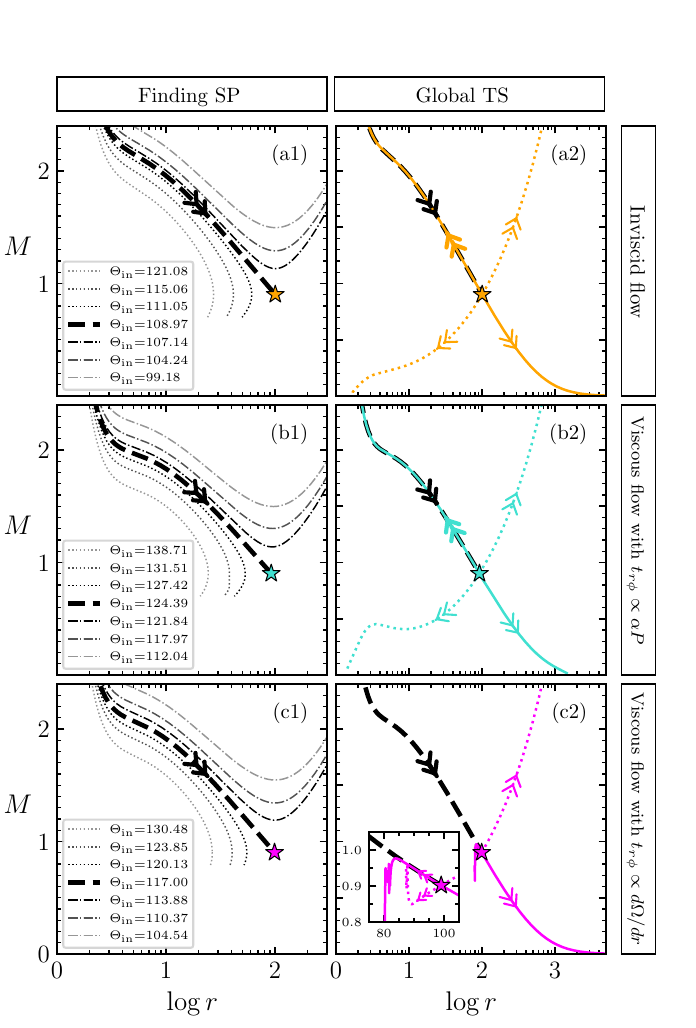}
\caption{\label{fig:findsp} Finding of SPs (left panel) and global transonic solutions (right panel) for different cases of dissipation (inviscid and two types of viscous prescription). The double arrows represent the direction of integration of the EoM. In the right panel, solid coloured curves represent the accretion solution and dotted {curves the} wind solution, with the SP marked using a star. Flow parameters used are, $E=1.001,~\lambda_0=1.45,~\alpha=0.03,~\beta=0.01,~\mdot=0.1\medd,~\mbh=10\msolar, ~\xi=1.0$.}
\end{figure}

\begin{enumerate}
\item {FLOW PARAMETERS}: Supply $E$, $\lambda_0$, $\alpha$, $\beta$, $\mdot$, $\mbh$ and $\xi$. Here, $\beta$ is the inverse of plasma beta parameter and is used to find the magnitude of stochastic magnetic field required for computing the synchrotron emission. This step defines or characterises the system. 
\item {BOUNDARY VALUES}, finding appropriate values of $\rin$, $\vin$, $\thetain$ and $\lambdain$: Select $\rin \rightarrow \rg =1.001$ to be the boundary and focus on the accretion branch. The reason for selecting this inner boundary is {that} asymptotically close to the central object, gravity is strong, and infall timescales are much shorter than any other timescales. Hence, matter does not get sufficient time to radiate. Thus, $E\simeq {\mathbb E}$ (see, Eq.~\ref{eq:gse}), and an algebraic form of energy is obtained. Simplifying it : $\vin=\sqrt{2[{\mathbb E}-h(\thetain)-\lambda_0^2/(2\rin^2)+1/{2(\rin-1)}]}$. 
Here, $\lambdain \approx\lambda_0$ is assumed, since their difference is generally less than $10^{-10}$\cite{bl03}. 
In this step, a guess value of $\thetain$ is supplied and $\vin$ is obtained from the above expression.
\item {IRM-SP-RV TECHNIQUE}: Using $\vin$, $\thetain$, $\lambdain$ (from step 2) and the flow parameters supplied (in step 1) 
at $\rin$, as boundary conditions (BCs), an integration technique is employed (preferably an explicit adaptive RK method) to solve the set of coupled differential equations \ie the EoM: $d\lambda/dr$ (Eq.~\ref{eq:dlamdr}), $dv/dr$ (Eq.~\ref{eq:dvdr1}) and $d\Theta/dr$ (Eq.~\ref{eq:flt}). It is to note that, in the IRM-SP scheme, which was also meant for viscous flows, the $d\lambda/dr$ was not solved since an algebraic expression of $\lambda$ was available due to the assumed viscous prescription of $\propto p$.

Coming back to the methodology, the solution obtained using the BCs, may not be transonic and could produce a supersonic branch (SB) or a multi-valued branch (MVB) solution. Hence, an iterative relaxation method\citep{press} is utilised, which has been extensively discussed in Paper 1, Section 3.1. Detailed diagnostics to identify the different branches were explained in the paper. For the relaxation of the solutions, the value of $\thetain$ is iterated, unless the solution satisfies the SP conditions at any $r=\rc$. This has been demonstrated in Fig.~\ref{fig:findsp}~(a1) which plots the Mach number ($M=v/\as$) vs $r$. For different values of $\thetain$s supplied, either SB (dashed-dot curves) or MVB (dotted curves) solutions are obtained. After a few $\thetain$ iterations, a solution is obtained (thick-dashed black curve) which passes through a SP (coloured star). The double arrows show the direction of integration: $\rin $ $\rightarrow$ $\rc$. 
Fig.~\ref{fig:findsp} (a1--a2) assumes no viscosity, (b1-b2) assumes viscous stress to be $\trpss$ (same as Paper 1), while the bottom panel (c1-c2) assumes $\trpr$, which is mainly investigated in the current work.  For the solutions represented in the upper and middle {panels}, the $d\lambda/dr$ equation is not solved and IRM-SP technique is utilised, while IRM-SP-RV is utilised for solutions plotted in panels c1-c2. The flow parameters assumed for these solutions are $E=1.001,~\lambda_0=1.45,~\alpha=0.03,~\beta=0.01,~\mdot=0.1\medd,~\mbh=10\msolar, ~\xi=1.0$.  

\item {CHECKING OF MSP}: To check whether there are any other SPs present (apart from the one obtained above) for the same set of flow parameters, the $\thetain$ used to obtain SP in step 3 is changed by a large factor, and the same IRM-SP-RV technique is followed. 
For the present case, in Fig.~\ref{fig:findsp}, only one SP is present.  However, if the parameters are changed, then MSPs may appear. 

\end{enumerate}

It is important to note here that, for locating SP since EoM are integrated in the forward direction, no specific integration method is mentioned in step 3. Use of any explicit integration method serves our purpose. These methods are fast and are hence used. As mentioned in BL03, the problem arises only during inwardly/negatively-directed integration (from SP), which has not yet been used in the methodology to find SPs. 

\subsection{Method to obtain global transonic solutions }

In the IRM-SP-RV scheme, $\rc$ and its corresponding flow variables $\vc$, $\thetac$ and $\lambdac$ are located. To obtain TS, the EoM are integrated inwards (negative $r$) and outwards (positive $r$) using the SP as the boundary. However, $\rc$ being a critical point,  the expression of $dv/dr|_{\rc}$ needs to be computed using L'Hopital's rule. This gives a quadratic expression in $dv/dr$ and two roots are obtained (see, Appendix \ref{app:A}). Negative $dv/dr$ produces the accretion solution, while the positive gives the wind solution. Fig.~\ref{fig:findsp} (a2), which is for the inviscid case, solid orange curve plots the solution with negative $dv/dr$ at $\rc$ and is the accretion solution: subsonic at infinity while supersonic near the surface; while dotted orange curve plots TS with positive $dv/dr$ and is the wind solution: subsonic near the surface and supersonic at infinity. The coloured double arrows indicate the direction of integration of the EoM from the SP. Similarly, for the viscous case with $\trpss$, the accretion (solid cyan) and wind solutions (dotted cyan) are presented in panel (b2). For the current set of flow parameters chosen, it is seen that irrespective of the nature of viscous dissipation, global TS for accretion as well as wind passes through a single SP (coloured star).

\subsubsection{The issue of integration}
The thick-dashed black curves in Fig.~\ref{fig:findsp} (a1, b1, c1) are overplotted on the global TS plotted in panels a2, b2 and c2, respectively. It is seen that the supersonic portion of the accretion solutions, in panels a2 and b2 (represented by thin, solid, coloured curves), exactly overlaps with the thick-dashed black curves. This suggests that the solution is retraced and validates the fact that finding of a solution should be independent of the direction of integration. However, in panel (c2), it is seen that when the viscous shear stress is $\trpr$, the inwardly-directed integration fails. This is exactly similar to the issue reported by BL03 (Fig. 5 of their paper). 
However, they were investigating only accretion solutions. Thus, they integrated outwards from the inner boundary to obtain the supersonic branch of the accretion solution, $\rin \rightarrow \rc$ (thick-dashed black curve in our case). Thereafter, they further integrated outwards to obtain the subsonic portion,  $\rc \rightarrow r_\infty$ (thin coloured solid curves of a2, b2). The integration direction was always in the outward direction, so the full global accretion solution could be obtained in BL03. 

The current work, however, deals with the unification of accretion and wind solutions as well as discusses flows with MSP, which might harbour shocks. Thus, this work investigates the issue presented in panel (c2) in detail. 
This situation comes because of the presence of a differential term $d\Omega/dr$  in the $d\lambda/dr$ equation (Eq.~\ref{eq:dlamdr}). This is a stiff equation. 
Thus, the traditional explicit integration methods fail to recover the solution. Even with higher order methods of RK or incorporating adaptive step-size refrains one from obtaining global transonic solutions. 

\subsubsection{Implicit-Explicit (ImEx) scheme}
Since the $d\lambda/dr$ equation is stiff, an implicit scheme is experimented upon and included in the picture of integration methodology \cite{l01,n00,t22},  unlike the sole use of traditional explicit schemes. It is seen and reported in this work, that an Implicit-Explicit (ImEx) scheme comes to the rescue  \cite{h96,h99} and could successfully as well as efficiently regenerate different parts of the transonic solution for accretion as well as wind. In the current work, the scheme developed follows the rule that an implicit scheme is used to solve the expression of $d\lambda/dr$; while using an explicit method for $dv/dr$ and $d\Theta/dr$. This combination is developed mainly because explicit methods are fast and thus, solving solely the stiff part with an implicit method is more efficient\cite{but01}. But, it is to mention here that implicit schemes are difficult to solve since it requires the values of the current as well as the next step to find the next step, $y_{n+1}=y_n+dy(y_{n+1},r_{n+1})/dr$ (representative for the simplest case). Thus, in order to avoid computational complexity, an implicit trapezoidal method is implemented \cite{nasa,a89}. For explicit integration, the simplest adaptive RK method is used, which involves combining Heun's method, which is of order 2, with the Euler method, which is order 1 \cite{s03,a98,but01,but03}. Their exact forms are expressed below in detail.
\begin{itemize}
\item General forms : The Butcher tables \cite{but01,but03} have the following form: 
\begin{equation}
\begin{array}{c|cccc}
c_1 & a_{11} & a_{12} & \ldots & a_{1s} \\
c_2 & a_{21} & a_{22} & \ldots & a_{2s} \\
\vdots & \vdots & \vdots & \ddots & \vdots \\
c_s & a_{s1} & a_{s2} & \ldots & a_{ss} \\
\hline & b_1 & b_2 & \ldots & b_s\\
 & b_1* & b_2* & \ldots & b_s*
\end{array}
\end{equation}
RK methods are represented by:
\begin{equation}
y_{n+1}=y_n+h\sum_{i=1}^s b_ik_i
\end{equation}
where, $$ k_i= f\left(x_n+c_i h, y_n+h\sum_{j=1}^{s-1} a_{ij} k_j\right)$$\\
Since an adaptive scheme is utilised, the lower-order derivative is given by:
\begin{equation}
y_{n+1}^*=y_n+h \sum_{i=1}^s b_i^* k_i
\end{equation}
where $k_i$s are the same as for the higher-order method. Then the error is given by: 
\begin{equation}
err_{n+1}=y_{n+1}-y_{n+1}^*=h \sum_{i=1}^s\left(b_i-b_i^*\right) k_i
\end{equation}
This $err$ value is used to compute the step-size \cite{press}.
\item{Current work}:  The Butcher tables used in the current paper are given in Table \ref{tab:bt} \cite{but01,but03}.
\end{itemize}
\begin{figure}
\includegraphics[scale=0.78,  trim={3.71cm 6.8cm 10cm 6.cm},clip]{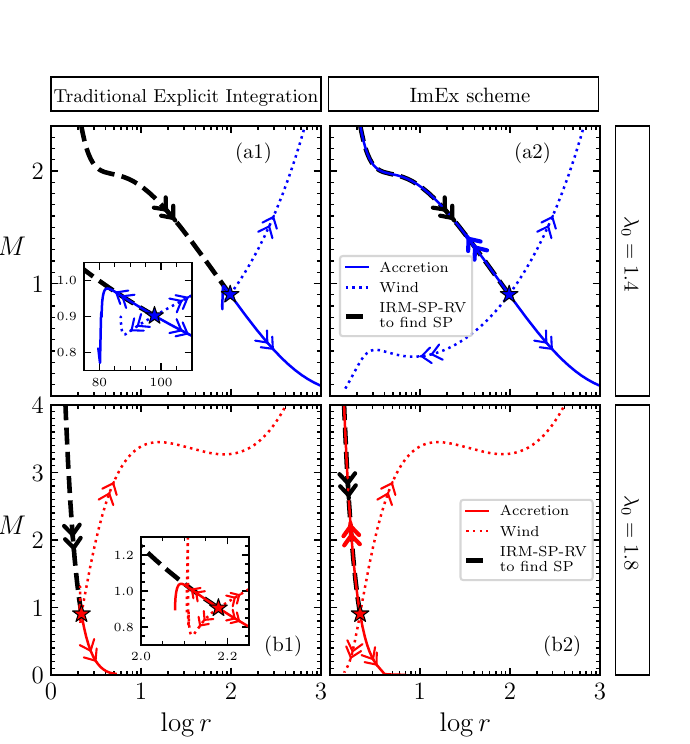}
\caption{\label{fig:imexsol} Failure of traditional explicit integration schemes to obtain TS (panels a1, b1) and employing ImEx scheme (panel a2, b2) successfully to obtain the global transonic accretion and wind solutions. Flow parameters used are same as in Fig.~\ref{fig:findsp} except for the $\lambda_0$ values.}
\end{figure}

\begin{table}[h]
\renewcommand\arraystretch{1.2}
\centering
\begin{subtable}[t]{0.17\textwidth}
\[
\begin{array}{l|l l}
    
0   &      \\ 
1   &   1   \\ \hline
    &  1/2 & 1/2 \\
    & 1 & 0\\
\end{array}
\]
\caption{Explicit Method}
\end{subtable}
    \hfil
\begin{subtable}[t]{0.3\textwidth}
\[
\begin{array}{l|l l}
    
0   &   0   & 0   \\ 
1   &   1/2 & 1/2   \\ \hline
    &  1/2 & 1/2 \\
    & 1 & 0\\
\end{array}
\]
\caption{Implicit Method}
\end{subtable}
\caption{Butcher's Tableaus}
\label{tab:bt}
\end{table}

This combination of Implicit-Explicit (ImEx) integration methods has been mainly used because the coefficients of step-size ($c_i$, left column of Butcher's table) are {exactly the} same. Thus, they point to the same $r$ after the coupled integration of the EoM is solved.

Utilising this methodology, plotted in Fig.~\ref{fig:imexsol} (a2, b2) are the global accretion and wind solutions. The same set of flow parameters are used as in Fig.~\ref{fig:findsp} except for the $\lambda_0$ values, which are taken to be 1.4 (Fig.~\ref{fig:imexsol}, a1, a2) and 1.8 (Fig.~\ref{fig:imexsol}, b1, b2). In the left panel, the traditional explicit integration schemes are employed to obtain the TS for accretion (solid curve) and wind (dotted curve). It is seen that the integration fails when it is inwardly directed (see, zoomed inset). The double arrows in all the panels represent the direction of integration. The thick-dashed black curve is the supersonic part of accretion solution obtained while finding the SP: $\rin \rightarrow \rc$ (coloured star) using IRM-SP-RV technique. The ImEx scheme when utilised could regenerate this supersonic accretion part, see panels a2 and b2. Thus, irrespective of the direction of integration, global transonic accretion and wind solutions could be obtained. For both the cases of flow parameters, the solution passes through a single SP: outer SP ($\rco$) for $\lambda_0=1.4$ and inner SP ($\rci$) for $\lambda_0=1.8$.

The main key point is that, only an implicit scheme can help obtain global \textit{wind} solutions. There is no other way one can find the subsonic branch of wind flows. Accretion solution can somehow be retraced using the scheme of BL03, by integrating only in the outward direction ($\rin \rightarrow \rc$ and then $\rc \rightarrow r_\infty$). But it is impossible to obtain global wind solutions. Unifying of accretion and wind solutions is necessary and has been worked on for a long time, but this serious issue has never been addressed.  
Thus, the ImEx scheme is necessary to be implemented in finding transonic global accretion and wind solutions, efficiently as well as accurately. 


\section{Results}
The IRM-SP-RV technique (to obtain SP(s)) coupled with the ImEx scheme (to obtain global solutions) is used to investigate accretion as well as wind flows around compact objects. In this section, the effect of variation of different parameters on the dynamics of the flow around compact objects is discussed. A global picture of flows encompassing a broad parameter space has been presented in the subsections below. 
Locating shocks in wind flows requires a different methodology, the IRM-SHOCK, as has been reported in Paper 1. The current work uses that and has been discussed in subsection \ref{sec:windshock}.

\subsection{Typical global transonic flows and their flow properties}
\begin{figure*}[]
\includegraphics[scale=0.7,  trim={1.cm 6.6cm 1cm 8.cm},clip]{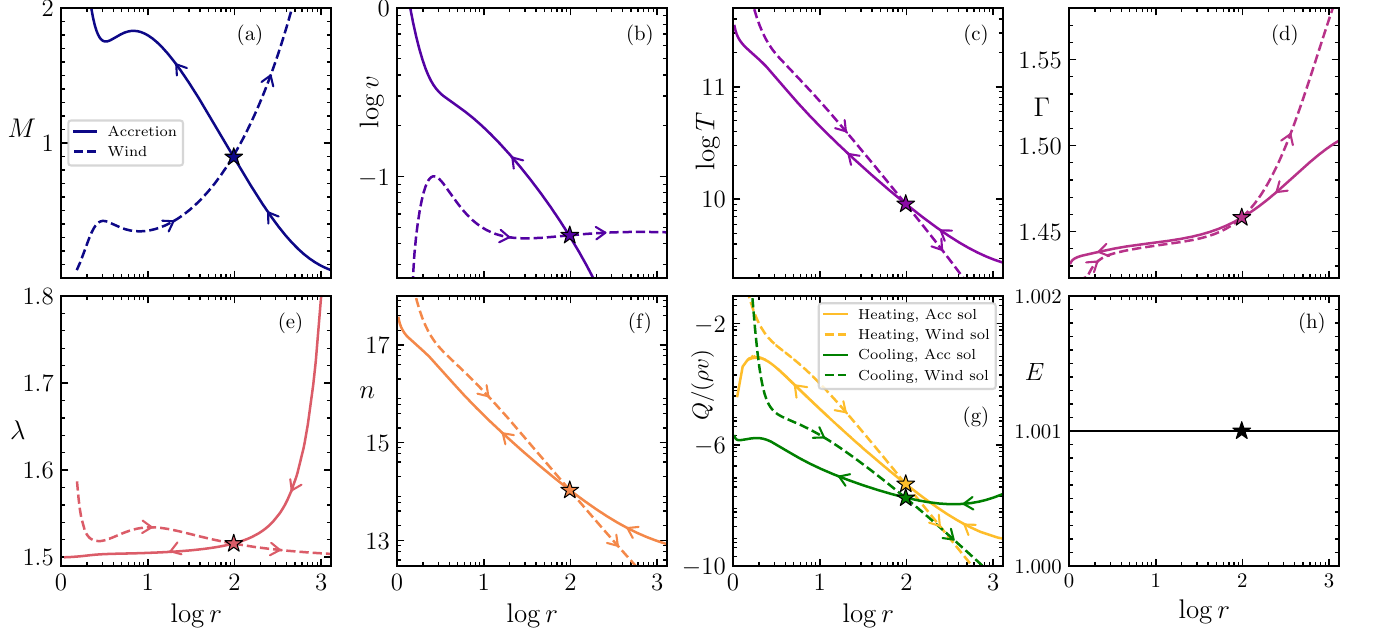}
\caption{\label{fig:sol} Typical global transonic solution for accretion (solid curves) and wind (dashed curves). The direction of flow is given by arrows. The flow parameters are: $E=1.001,~\lambda_0=1.5,~\alpha=0.01,~\beta=0.01$, $\mdot=0.1\medd$, $\mbh=10\msolar$, $\xi=1.0$.}
\end{figure*}
In Fig.~\ref{fig:sol}, a typical TS is plotted along with its flow variables, obtained for flow parameters $E=1.001,~\lambda_0=1.5,~\alpha=0.01,~\beta=0.01,~\mdot=0.1\medd,~\mbh=10\msolar, ~\xi=1.0$. Both the accretion and wind solutions for the given set of flow parameters harbour a single outer SP at $\rco=98.255$ (coloured star). The accretion solution is plotted using solid curve while the wind solution in dashed curve.  Since both the accretion and wind solution is connected with the inner boundary, IRM-SP-RV technique was sufficient to locate the SP. The single arrows represent the direction of flow, unlike in the previous section figures, where double arrows were used to represent the direction of integration of the EoM. In panel (a) the solutions are represented using $M$ and in (b) the flow velocities are plotted. It is seen that for the accretion solution (solid curve), the flow starts from very low velocities and crosses the speed of light near the surface.  This is mainly because of the efficient conversion of gravitational potential energy into kinetic energy of the matter.  The wind solution (dashed curve), however, shows an opposite nature. It starts with very negligible velocities and gain very high values just after the ejection. At later stages of the flow, matter attains some constant velocity. 
Panel (c) plots the temperature ($T$) and (d) the adiabatic index ($\Gamma$). It is seen that the accretion flow is non-relativistic ($\Gamma \sim 5/3$) with very low $T$ far away. As the matter is accreted, it is compressed and becomes mildly relativistic, reaching higher temperatures near the surface of the compact object. 
For the wind case, the matter is hot near the surface, and as it is moving away, it is expanding, leading to a lowering of the temperature and hence increase in $\Gamma$ values. From panel (e) it is evident that the viscosity has effectively transferred angular momentum outwards for the accretion case. In the case of winds, the matter is seen to be rotating much more near the surface while it is less rotating or more radially flowing far away from the central object. The number density in units of cm$^{-3}$ is plotted in panel (f). As can be seen from the figure, accretion is a converging flow while wind is a diverging flow. Panel (g) plots the dissipation present in the system. The cumulative heating (yellow curves) and cooling (green curves) are plotted for accretion (solid) and wind (dashed) flows. For the present set of flow parameters, heating dominates most of the region of the accretion as well as wind flows. In case of accretion, the infall timescales are extremely short just near the central object, because of large infall velocities. In these regions, the magnitude of dissipation decreases. The matter does not get enough time to radiate as well as to become viscously heated. This effect is seen from the dip in their curves (solid yellow and green) just near the surface.  Panel (h) reiterates the fact that the generalised Bernoulli constant (Eq.~\ref{eq:bernoullidisc}) is a constant of motion throughout the flow, even in the presence of dissipation. 

\subsection{Shocks in accretion and wind flows}

In certain regions of parameter space, accretion and wind flows may have MSP and hence can harbour shocks. The methodology to find MSPs in accretion and wind flows differs. For all topologies of accretion flows, solutions passing through both the SPs are connected to the inner boundary, while for wind flows it is not necessary that both the solutions are connected. Thus, finding of MSPs or shocks in accretion flows are comparatively easier than in wind flows. It is important to remember that the global solution would pass through both the SPs only if the solution harbours a shock. For accretion flows, matter starts from infinity at subsonic velocities $\rightarrow$ gets accreted and becomes supersonic after passing through $\rco$ $\rightarrow$ encounters a shock transition ($\rsh$) and becomes subsonic $\rightarrow$ again becomes supersonic after being accreted and passing through $\rci$ $\rightarrow$ plunges into the horizon.  For wind flows, the direction is just the opposite, and matter starts from a region near the central object with subsonic velocities, becomes supersonic after passing through $\rci$ $\rightarrow$ encounters a shock and becomes subsonic $\rightarrow$ passes through $\rco$ and becomes supersonic. In accretion flows, the solution passing through $\rci$ is non-global while that passing through $\rco$ is always global. By global, it is meant that the flow is connected from infinity to the inner boundary of the compact object. In the absence of shocks, the global accretion flow would follow the solution passing through $\rco$. It is important to note here, that in certain cases, solutions passing through both $\rci$ and $\rco$ could be global. The parameter space where these types of global solutions exist is extremely small. In such a case, the solution with higher entropy is selected, since nature would prefer this solution, following the second law of thermodynamics. In case of wind flows, the MSP solutions are complicated and either $\rci$ or $\rco$ could be non-global. In these cases, the other SP is global. These types of systems have been elaborately discussed in the upcoming sections. 

In the subsections below, the first part discusses shocks in accretion flows and the second part discusses wind flows.

\subsubsection{Accretion flows with shocks}
\begin{figure}[]
\includegraphics[scale=0.75,  trim={1.9cm 11.6cm 6.8cm 8.cm},clip]{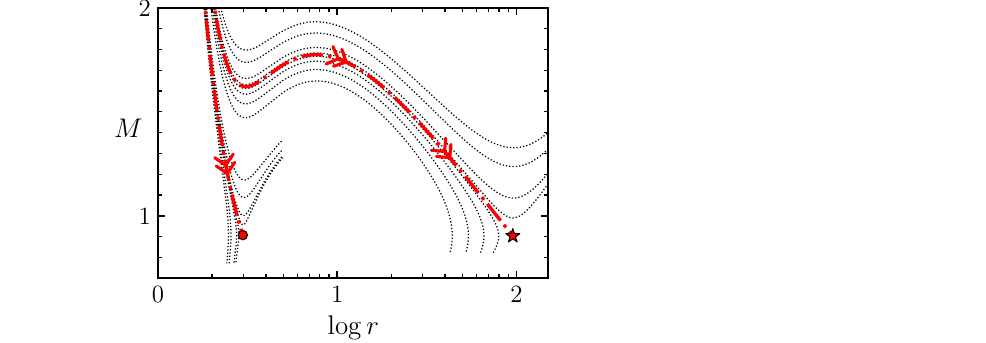}
\caption{\label{fig:findspboth} Finding multiple sonic points using IRM-SP-RV methodology. Thick dashed-dotted red curves denote the solutions passing through $\rco$ (red star) and $\rci$ (red circle). The black dotted curves are either MVB or SB solutions and hence are unphysical. 
The flow parameters are: $E=1.001$, $\lambda_0=1.5$, $\alpha=0.03$, $\beta=0.01$, $\mdot=0.1\medd$, $\mbh=10\msolar$, $\xi=1.0$.}
\end{figure}
\begin{figure}[]
\includegraphics[scale=0.7,  trim={3.7cm 5.6cm 11.5cm 7.2cm},clip]{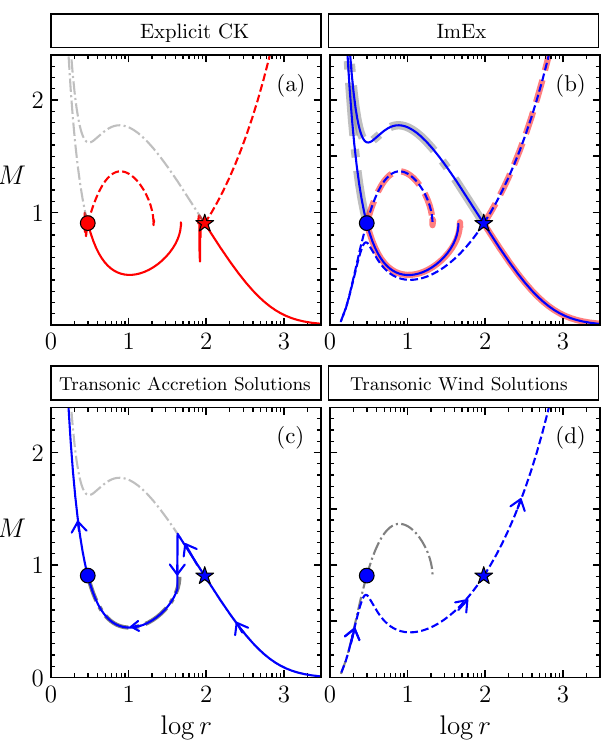}
\caption{\label{fig:shocksol} Typical global transonic solutions for accretion (solid curves) and wind (dashed curves). The direction of flow is given by arrows. The flow parameters are same as in Fig.~\ref{fig:findspboth}.}
\end{figure}

\begin{figure*}[]
\includegraphics[scale=0.7,  trim={0.9cm 6.6cm 1.5cm 8.cm},clip]{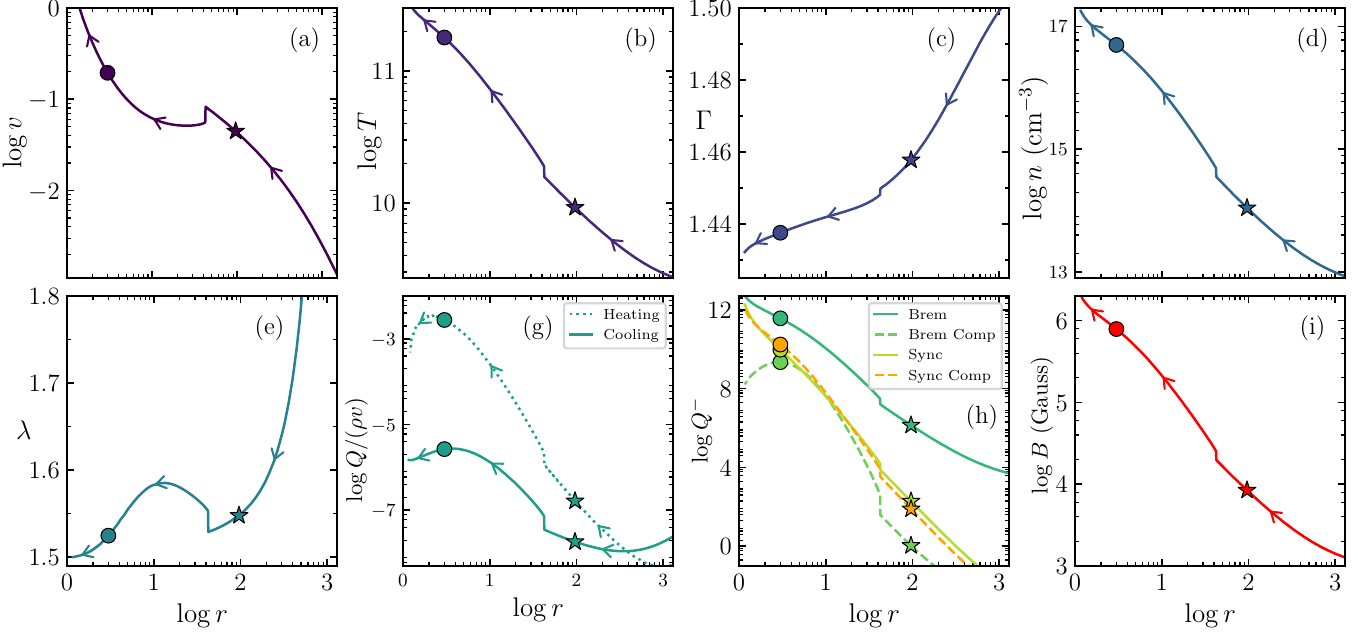}
\caption{\label{fig:shockpar} Flow variables in accreting systems harbouring shock. Couloured star represents $\rco$ and coloured circle represents $\rci$. The flow parameters are same as in Fig.~\ref{fig:findspboth}.}
\end{figure*}
As have been mentioned before in detail, accretion flows passing through both $\rci$ and $\rco$ are connected to the inner boundary. The solution passing through $\rci$ is non-global (NG), and $\rco$ is global (G). Since both these G and NG solutions are connected to the inner boundary, IRM-SP-RV technique is enough to obtain the location of the SPs. Plotted in Fig.~\ref{fig:findspboth} is the relaxation method to obtain MSP. The flow parameters used are $E=1.001,~\lambda_0=1.5,~\alpha=0.03,~\beta=0.01$, $\mdot=0.1\medd$, $\mbh=10\msolar$, $\xi=1.0$. The double arrows indicate the direction of integration. For  $\thetain=137.943$, $\rco$ is obtained (red star) and $\thetain=120.406$ gives the location of $\rci$ (red circle). The thick dashed-dotted curve represents the supersonic part of transonic solution obtained through IRM-SP-RV method. Other black dotted curves represent either MVB or SB, which are unphysical. After obtaining the location of both the SPs, the EoMs are integrated outwards and inwards to obtain accretion and wind solutions. Presented in Fig.~\ref{fig:shocksol} panel (a) the solution obtained using traditional explicit integration methods, in our case the CK method. It is seen that integration fails when it is inwardly directed. However, outwardly directed integration was successful, and one could obtain the subsonic branch of accretion flow (solid red curve) and the supersonic branch of wind flow (dashed red curve). $\rco$ is marked using red star and $\rci$ using red circle. The grey dashed-dotted curve represents the part of the solution obtained while finding SPs \ie the red dashed-dotted curve of Fig.~\ref{fig:findspboth}.

Panel (b) plots the complete solution obtained by integrating the EoM using ImEx scheme developed in this paper, as described in the methodology section. All the mathematically obtained accretion solutions are plotted using blue solid curves, and the wind solutions using blue dashed curves. They retrace the branches of solutions obtained in panel (a) using explicit CK scheme. They have been overplotted with less transparency in panel (b). This validates the fact that the new methodology developed is appropriate and helps one to obtain solutions of stiff EoM. To investigate these solutions better, the accretion and wind solutions have been separately plotted in panels (c) and (d).
It is seen that both accretion and wind solutions harbour MSPs (blue circle and star) and their solutions are connected to the inner boundary.  The solution passing through $\rci$ is NG (blue circle) and through $\rco$ is G (blue star) in case of accretion flows. This is just the reverse of what is seen in case of wind flows (see panel d). The direction of flow is represented using coloured single arrows.

On investigation, it is seen that the wind solution cannot harbour a shock because the solution passing through $\rci$ has a higher entropy than the solution passing through $\rco$, \ie $\mdotrci>\mdotrco$. Thus, the solution cannot jump from a higher entropy solution, represented using dashed-dotted grey curve to a lower entropy solution (dashed blue curve). However, in the present case, the higher entropy solution is NG. Thus, the global transonic wind solution would pass through $\rco$ as seen in panel (d). For accretion solution the direction of flow is opposite to wind. The lower entropy global solution passing through $\rco$ can jump to a higher entropy NG solution passing through $\rci$ provided the shock conditions (see, section \ref{sec:shocks}) are satisfied. In the present set of flow parameters, it is found that a shock is present at $\rsh=42.630$, where the mass, momentum and energy fluxes match. This suggests that the global solution would pass through $\rco\rightarrow\rsh\rightarrow \rci$. This is plotted in Fig.~\ref{fig:shocksol}c, blue solid curve. 
If the shock conditions were not satisfied then the global solution would have passed through $\rco$ (blue star) and the solution is represented using dashed-dotted grey curve. 

Accretion flows harbouring shocks are expected to have shock signatures in the different flow variables. In Fig.~\ref{fig:shockpar}, (a) velocity, (b) temperature ($T$) in Kelvin (c) adiabatic index ($\Gamma$), (d) number density in units of ${\rm cm}^{-3}$ and (e) $\lambda$ is plotted as a function of radius. At the shock location, velocity decreases while temperature, number density and angular momentum increase due to compression. Since $T$ rises,  the flow becomes more relativistic, resulting in the decrease of $\Gamma$. In the presence of shocks, the decrease in radial velocity is accompanied by an increase in azimuthal component of velocity, leading to an increase in angular momentum (see panel e). Panel (g) plots the total heating (dotted green curve) and total cooling (solid green curve) in units of $\rho v$. Thus, these quantities are dimensionless. The flow is heating-dominated. In panel (h) different cooling processes in units of ergs/cm$^{3}$/s are plotted. Bremsstrahlung and synchrotron are represented using solid curves, while their respective Comptonizations using dashed curves. It is seen that bremsstrahlung dominates the cooling throughout the flow. Generally, bremsstrahlung photons are hard and hence it is reflected through their mild Comptonized component. On the other hand, synchrotron photons cover a broad range in the electromagnetic spectrum. Thus, their photons are Comptonized more than in the bremsstrahlung case, especially near the central object. The magnetic field is plotted in panel (i), which is obtained assuming an equipartition with gas pressure.

\subsubsection{Wind flows with shocks}\
\label{sec:windshock}
\begin{figure}[]
\includegraphics[scale=0.55,  trim={3.6cm 11.5cm 6.8cm 7.2cm},clip]{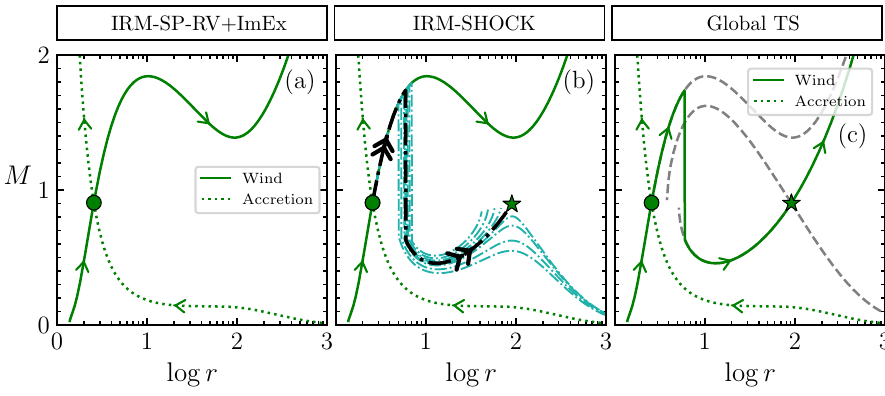}
\caption{\label{fig:irmshock} Typical global transonic solutions for accretion (dotted curve) and wind (solid curve). $\rci$ is represented using green circle and the direction of flow is given by arrows. The flow parameters are: $E=1.001,~\lambda_0=1.6,~\alpha=0.03,~\beta=0.01$, $\mdot=0.1\medd$, $\mbh=10\msolar$, $\xi=1.0$.}
\end{figure}

Unlike accretion flows, MSP for winds are difficult to obtain. While the solution passing through $\rci$ is always connected to the inner boundary the solution passing through $\rco$ is not necessarily so. 
In the section before, it was seen that if the wind solution passing through $\rco$ was global, nature prefers this solution, considering its higher entropy.
For the cases where $\rco$ is NG and not connected (from infinity) to the inner boundary, it is important to remember that locating this SP is necessary only when shocks are formed. This is because of the fact that a global solution would pass through both the SPs when it harbours a shock. This led to the development of IRM-SHOCK technique, elaborately discussed in Paper 1. If a wind flow harbours a shock, then only it is possible to locate $\rco$ and the global TS would pass through both the SPs via a shock transition. Otherwise, the wind solution would pass through either $\rci$ or $\rco$ whichever is global and has a higher entropy.

The IRM-SHOCK technique is briefly discussed below:
\begin{enumerate}
\item Use IRM-SP-RV to find $\rci$ for a given set of flow parameters: $E$, $\lambda_0$, $\alpha$, $\beta$, $\mdot$, $\mbh$ and $\xi$.
\item Obtain the global transonic wind solution passing through $\rci$ using ImEx scheme.
\item At each point of the supersonic branch of wind solution, the shock conditions are utilised to find post-shock values. Using these values, we assume a shock jump at that location, and the EoM are integrated further outwards. The post-shock solution thus obtained need not be transonic.
\item Step 3 is iterated \ie relaxation method is applied by iterating on $r$ until the post-shock solution passes through a SP. In this way $\rco$ is located.
\end{enumerate}
It could be possible that the post-shock solution does not pass through {an} SP, even after multiple iterations of $r$. This would suggest that the $\rco$ does not exist, and the global transonic solution for wind would pass through $\rci$.

\begin{figure}[]
\includegraphics[scale=0.72,  trim={1.cm 6.6cm 12.8cm 7.9cm},clip]{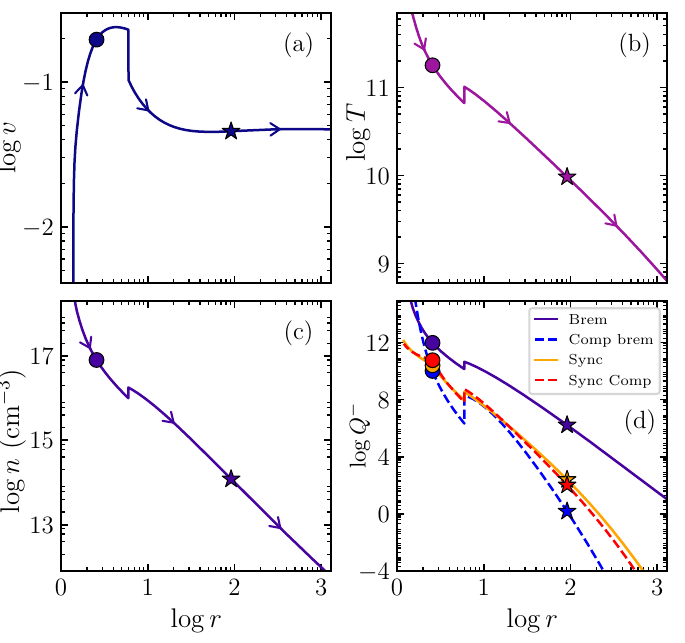}
\caption{\label{fig:windpar} Flow variables associated with the shocked wind solution of Fig.~\ref{fig:irmshock}c (green solid curve). The direction of flow is given by arrows. }
\end{figure}
\begin{figure*}[ht]
\includegraphics[scale=0.8,  trim={0.cm 3.9cm 3.cm 6.5cm},clip]{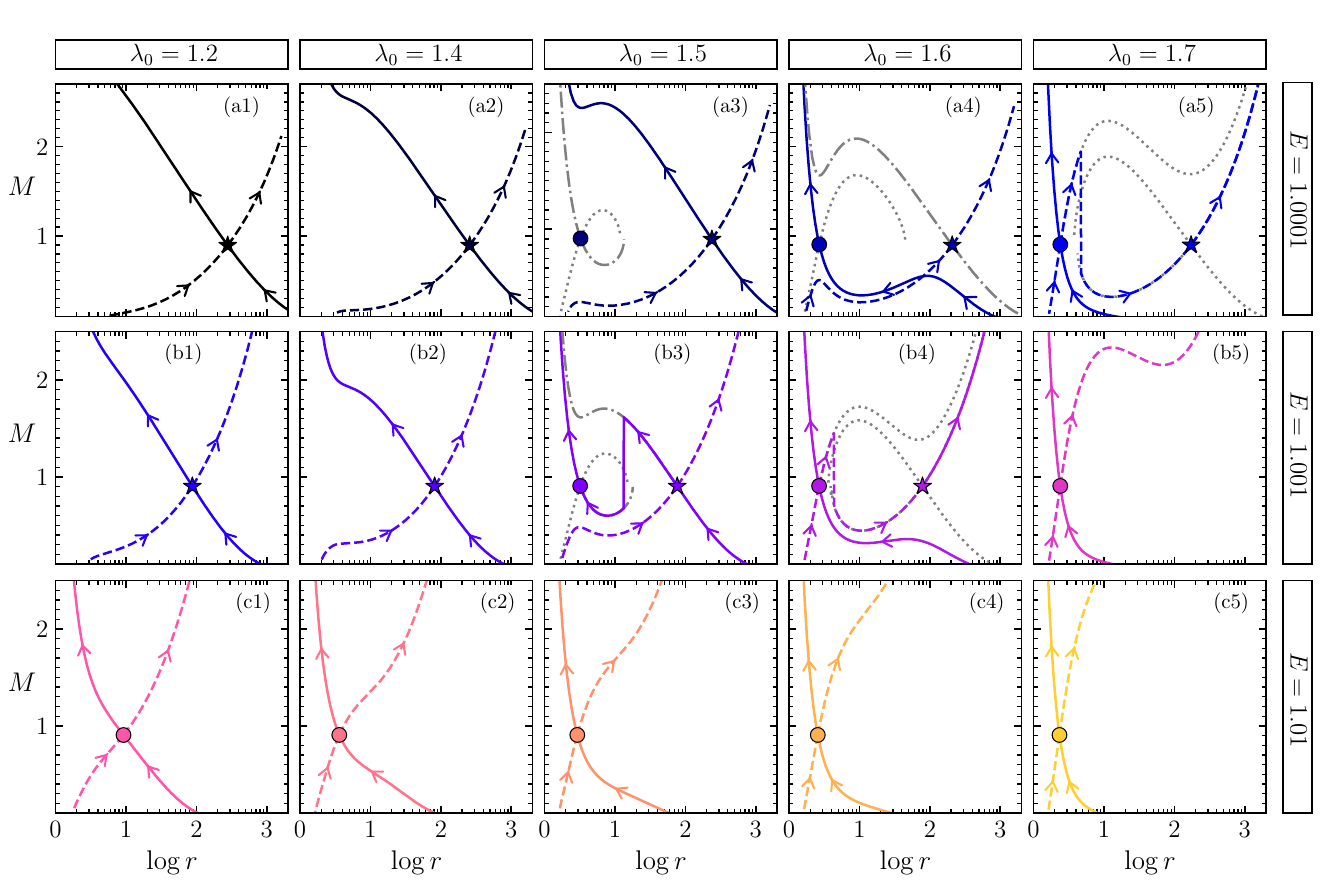}
\caption{\label{fig:ange} Typical global transonic solutions of accretion (solid curves) and wind (dashed curves) for variation of $\lambda_0$ with $E$ is plotted. The flow parameters are: $E=1.001,~\alpha=0.01,~\beta=0.01$, $\mdot=0.1\medd$, $\mbh=10\msolar$, $\xi=1.0$.}
\end{figure*}

\begin{figure*}[]
\includegraphics[scale=0.7,  trim={0.1cm 2.cm 3.6cm 2.2cm},clip]{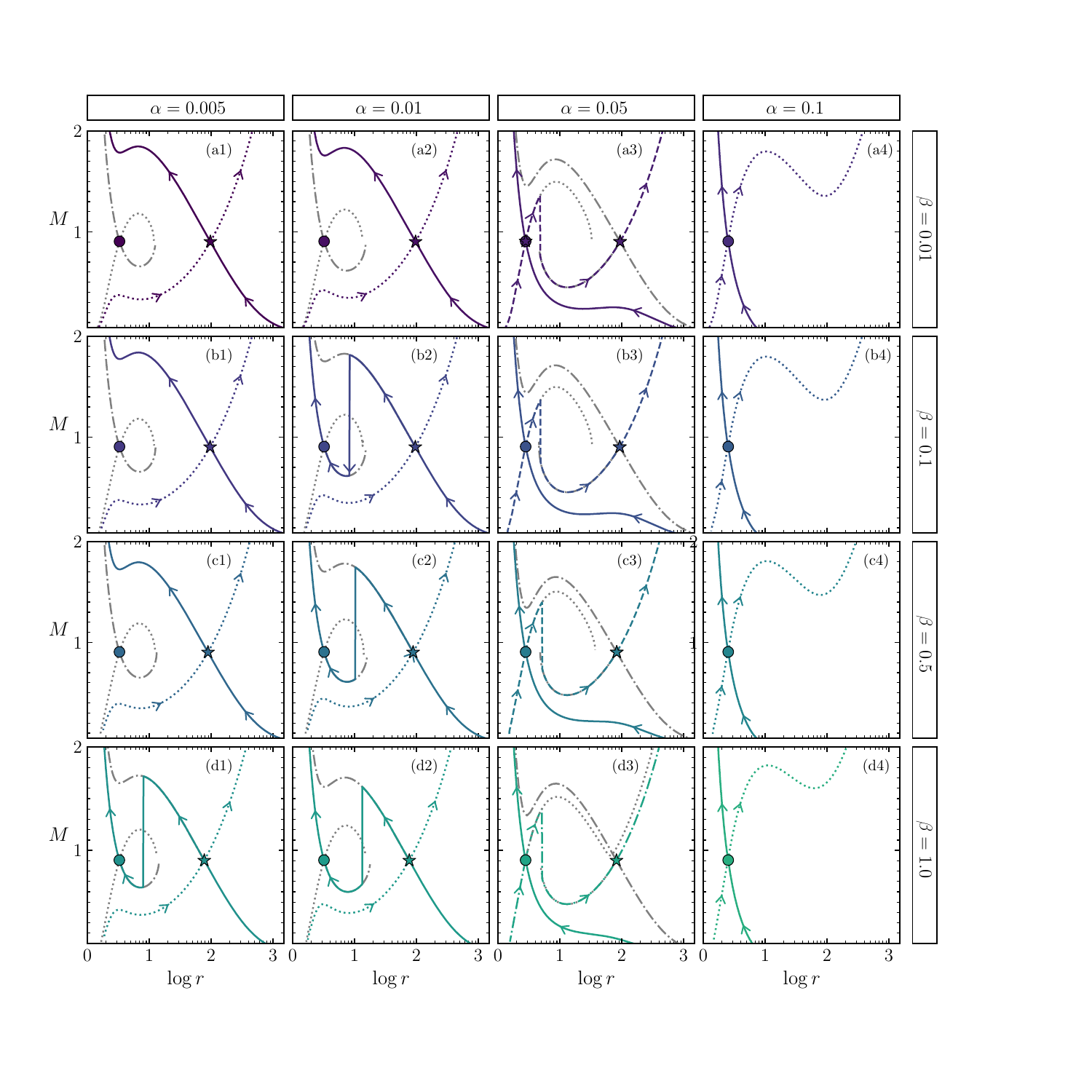}
\caption{\label{fig:alphabeta} Typical global transonic solution for accretion (solid curves) and wind (dotted curves). The direction of flow is given by arrows. The flow parameters are: $E=1.001,~\lambda_0=1.5$, $\mdot=0.1\medd$, $\mbh=10\msolar$, $\xi=1.0$.}
\end{figure*}

The IRM-SHOCK technique is explained in Fig.~\ref{fig:irmshock}. The flow parameters used are $E=1.001,~\lambda_0=1.5,~\alpha=0.01,~\beta=0.01$, $\mdot=0.1\medd$, $\mbh=10\msolar$, $\xi=1.0$. Panel (a) presents the global transonic wind (solid green) and accretion (dotted green) solution. Step 1 and 2 is utilised \ie IRM-SP-RV technique to find $\rci$ and the global transonic solutions using ImEx scheme. Panel (b) represents {steps} 3 and 4, where at every $r$ in the supersonic wind branch, shock jump is initiated. Thereafter, the EoM are integrated to obtain a solution and checked for its transonicity. For the present set of flow parameters, an $\rco$ exists, represented by a green star. In panel (c) the EoM are integrated outwards from $\rco$ to find the complete global transonic wind solution. The solutions plotted in grey are unphysical branches. For the present set of flow parameters, wind solution passes through both the SPs via a shock transition (solid green curve). It is important to note that through $\rco$ a mathematical accretion branch also exists, represented by dashed grey curve. However, this solution passing through $\rco$ is NG and has a higher entropy. Thus, an accretion shock is unable to form and the global accretion solution would pass through $\rci$.

In Fig.~\ref{fig:windpar} plotted are few of the flow variables associated with the shocked wind flow presented in Fig.~\ref{fig:irmshock}. All these flow variables have shock signatures which are similar to that observed in accretion flows (see, Fig.~\ref{fig:shockpar}). The direction of flow is just opposite to that of accretion and after the shock, velocity decreases while $T$ and $n$ increases similar to what has been observed in accretion flows. The main difference is that the variation of $T$ and $n$ values are much larger, the change in these flow variables with $r$ is rather broad. The emissivity in units of ergs/cm$^3$/s (panel d) shows an overall similar pattern as have been observed in accretion flows and bremsstrahlung emission dominates throughout the flow. This is due to the fact that $\beta$ values in both the accretion and wind case is low and is equal to 0.01 while accretion rate is $\mdot=0.1$. Thus, synchrotron emission is weak. A higher accretion rate is accompanied by an increase in number density of the flow. Since, $\qbr \propto n^2$, the bremsstrahlung emission dominates throughout the flow.

\subsection{Flow properties with variation in energy and angular momentum}

Bulk angular momentum at the horizon ($\lambda_0$) and Bernoulli constant are important parameters which define the flow. In Fig.~\ref{fig:ange}, the variation of both these parameters are plotted. $\lambda_0$ is increased from left to right (panels 1-5) while $E$ is increased from top to bottom (panels a-c). The dashed curves represent wind flows while solid curves represent accretion solutions. The direction of flow is represented by arrows.  $\rci$ is represented using coloured circle while $\rco$ using coloured star. The grey curves represent solution which are either non-global or not selected because of less entropy.   It is seen that as angular momentum is increased keeping $E$ constant,  both the accretion and wind solutions transition from flowing through outer sonic point to inner sonic point. This is because as the angular momentum is increased, the flow gets enough time to interact, increasing the efficiency of dissipation processes. Specifically, it increases the amount of viscous dissipation, making the system hotter. A flow encounters a SP when the velocity of the flow equals the sound speed. For a hotter flow, the sound speed rises. Thus, the SP will now be formed at a place where the velocity is larger. This condition is true near the central object. Hence, with increasing value of angular momentum the SP through which global solution passes shifts from $\rco$ to $\rci$. It is important to see that in between the two extreme limits of $\lambda_0$, MSP forms in certain combination of $\lambda_0-E$ parameter space. Inside this MSP regime, certain solutions harbour shocks as well. Accretion shocks are seen to be formed in panel b3 while wind shocks are formed in panels a5 and b4. The latter is obtained using IRM-SHOCK technique. 

Increasing $E$ means the thermal energy at infinity is increased. The system becomes hotter with increasing $E$ and the same argument holds, as have been discussed before, for hotter flows. Initially solution passes through $\rco$ (see, panel a3) and thereafter with increasing $E$, MSP starts forming, with few harbouring shocks (see panel b3). For higher energies, it is seen that only one SP forms. This is because, the temperature of the system is so high, that the transonicity condition could only be satisfied near the central object. 

\subsection{Flow properties with variation in dissipation parameters $\alpha$ and $\beta$}
In this section, the variation of different values of $\alpha$ and $\beta$ on the topology of solutions is discussed.  In Fig.~\ref{fig:alphabeta}, $\alpha$ increases from left to right (panels 1-4) while $\beta$ increases from top to bottom (panels a-d), values of which are mentioned inset. The effect of $\alpha$ is {primarily to} increase the heating while that of $\beta$ is to increase the cooling through increasing of synchrotron emission. 
It is seen that as $\alpha$ is increased, keeping $\beta$ constant, the global accretion solutions which initially passed through $\rco$ passes through $\rci$. In between this transition, there are solutions which harbour shocks. As cooling is increased by increasing $\beta$, shocks are formed even for lower values of $\alpha$. A higher dissipation suggests a hotter flow at infinity. 
Thus, even for lower values of $\alpha$ shocks were formed. {A similar} trend is observed for wind flows as well.  A parameter scan reveals that interplay between heating and cooling processes leads to the rise of different types of solutions. 

\section{Concluding Remarks}
Current work is based on the problem identified by \fullcite{bl03}. It was argued that any inward-directed integration fails, although an outwardly directed integration is possible. These integrations were done using SP as a boundary and were required to obtain global transonic solutions. 
The issue arises when the flows around BHs are considered to be viscous and hence, dissipative. Paper 1 (\fullcite{sk25}) also dealt with these types of systems, but the issue never arose. This is mainly due to the fact that an approximate viscous treatment was used, which reduced the angular momentum equation to an algebraic form. This form was just a function of the flow variables. BL03, however, used a more realistic viscous treatment and hence, integration of $d\lambda/dr$ equation was needed along with the integration of the other EoM. 
The problem of integration pointed out in BL03 and in the current work is just because of the presence of stiff terms in the angular momentum equation.

The failure of integration depending on direction is indeed a serious issue.  BL03 skirted out the problem since they dealt with only accretion flows. The inner boundary, which is close to the event horizon, was utilised in their work. They integrated outwards from this boundary to find the SP through a relaxation technique. After finding the SP they further integrated outwards until they reached a point which is very far from the central object and has very less velocity. In this process of finding the accretion solution, the integration of EoM was always outwardly directed throughout. The current work, however, is not constrained to obtaining only accretion solutions but wind solutions as well. 
Velocity of accretion flow at the inner boundary is approximately the speed of light, but for wind flows it is unknown and could be as less as $10^{-5}c$ or less than that. Thus, one cannot use the same concept of integration from the inner boundary to find wind solutions. The generic method which has been used in literature, is to find the SP for a given set of flow parameters and then integrate the EoM inwards and outwards, using SP as boundary. The SP has a $0/0$ form and thus the differential equation of velocity reduces to a quadratic equation, which would result in two values of slope: one for accretion and another for wind. While the outwardly directed integration is found to be successful, the inwardly is not. This has been reported by BL03, as well as it has been elaborately discussed in the current work. This work has pointed out the issue and the major repercussions {of} finding a proper transonic solution. The wind solution can never be obtained, if the problem persists. 

The current work points out that traditional integration schemes need to be modified to curb the problem. While adaptive higher-order explicit methods are efficient and popular, they fail to solve the issue. 
An Implicit-Explicit (ImEx) scheme has been developed where the implicit scheme is used to integrate the angular momentum equation and an explicit scheme to solve the other EoM ($dv/dr$ and $d\Theta/dr$). This scheme could successfully produce all global accretion and wind solutions. 
An age-old problem of finding global transonic solutions due to a mathematical restriction has been solved for the first time.
This methodology has been used to obtain all classes of global solutions, MSP regime as well as shocks. A broad parameter space has been scanned, and it has been found that it could successfully produce all possible solutions.

In an upcoming paper, the different spectral signatures of accretion and wind will be investigated. This is important in order to understand the observed luminosities and transient phenomena going on in these systems. Shocks and QPO's are shown to affect the luminosities for a brief period of time. These areas would be investigated in the upcoming paper.

\begin{acknowledgments}
The author wants to thank Prof. Igor Kulikov for his helpful discussion.
\end{acknowledgments}

\appendix

\section{Finding the value of $dv/dr|_{\rm c}$ at the sonic point}
\label{app:A}
The expression of ${dv}/{dr}$ is given by Eq.~$\ref{eq:dvdr2}$. At the SP, it has a $0/0$ form. Thus, the derivative of velocity at SP is obtained using L'Hopital's rule and can be written as:

\begin{eqnarray}
\cfrac{dv}{dr}\bigg \vert_c&=&\cfrac{\cfrac{d{\cal N}(r,v,\Theta,\lambda)}{dr}}{\cfrac{d{\cal D}(v,\Theta,\lambda)}{dr}} \nonumber\\
\cfrac{dv}{dr}&=&\cfrac{\cfrac{\partial{\cal N}}{\partial r}+\cfrac{\partial{\cal N}}{\partial v}\cfrac{dv}{dr}+\cfrac{\partial{\cal N}}{\partial \Theta}\cfrac{d\Theta}{dr}+\cfrac{\partial{\cal N}}{\partial \lambda}\cfrac{d\lambda}{dr}}{\cfrac{\partial{\cal D}}{\partial v}\cfrac{dv}{dr}+\cfrac{\partial{\cal D}}{\partial \Theta}\cfrac{d\Theta}{dr}+\cfrac{\partial{\cal D}}{\partial \lambda}\cfrac{d\lambda}{dr}} \label{eq:app1}
\end{eqnarray}
In the above equation, everything is computed at the SP and the subscript $c$ has been removed for clarity.
The expression of $d\Theta/dr$ in Eq.~\ref{eq:flt} can be written as $C_1+C_2( dv/dr)$ where $C_2$ are the coefficients of $dv/dr$ and $C_1$ has rest of the terms. Substituting this in equation above (Eq.~\ref{eq:app1}) and simplifying it:

\begin{equation}
A_1\left(\dfrac{dv}{dr}\right)^2+A_2\left(\dfrac{dv}{dr}\right)+A_3=0
\end{equation}

where, 
\begin{eqnarray}
A_1&=&\left(\dfrac{\partial \cd}{\partial v}+\dfrac{\partial \cd}{\partial \Theta} C_2 \right)\nonumber\\
A_2&=&\dfrac{\partial \cd}{\partial \Theta}C_1-\dfrac{\partial \cn}{\partial v}-\dfrac{\partial \cn}{\partial \Theta}C_2,\nonumber \\
A_3&=&-\left({\dfrac{\partial \cn}{\partial r}+\dfrac{\partial \cn}{\partial \Theta} C_1+\dfrac{\partial \cn}{\partial \lambda}\dfrac{d\lambda}{d r}}\right)\nonumber \\
\end{eqnarray}
Thus, $dv/dr$  can now be {expressed} as a quadratic equation, the solution of which is given by: 
\begin{equation}
\cfrac{dv}{dr}\bigg\rvert_c=\dfrac{-A_2\pm \sqrt{A_2^2-4A_1 A_3}}{2A_1}
\end{equation}
The above equation helps in computing the value of $dv/dr$ at the SP.

\bibliography{ms}

\end{document}